\documentclass[pra,superscriptaddress,showpacs,papersize=a4paper,twocolumn,floatfix]
{revtex4-1}%
\usepackage[pdftex]{graphicx}
\usepackage{bm}
\usepackage{subcaption}
\usepackage{etoolbox}
\usepackage[usenames,dvipsnames]{color}
\usepackage{mathtools}
\usepackage{amsmath}%
\usepackage{amsfonts}%
\usepackage{amssymb}

\DeclarePairedDelimiterX\braket[2]{\langle}{\rangle}{#1 \delimsize\vert #2}

\newcommand{\bg}{ \begin{gather} }

\newcommand{\eg}{\end{gather}}
\newcommand{\be}{ \begin{equation} }
\newcommand{\ee}{\end{equation}}
\newcommand{\bea}{ \begin{eqnarray} }
\newcommand{\eea}{\end{eqnarray}}

\def\eps{\epsilon}
\def\tr{\textrm{tr}}

\newcommand{\sgn}{\mathop{\rm sgn}}

\AtEndEnvironment{thebibliography}{
}

\begin{document}

\title{SYK model with quadratic perturbations: the route to a non-Fermi-liquid.}
\author{A. V. Lunkin}
\affiliation{Skolkovo Institute of Science and Technology, 143026 Skolkovo, Russia}
\affiliation{ L. D. Landau Institute for Theoretical Physics, Kosygin str.2, Moscow 119334, Russia}
\affiliation{ Condensed-Matter Physics Laboratory, National Research University "Higher
	School of Economics", 101000 Moscow, Russia }
\author{K. S. Tikhonov}
\affiliation{ L. D. Landau Institute for Theoretical Physics, Kosygin str.2, Moscow 119334, Russia}
\affiliation{ Condensed-Matter Physics Laboratory, National Research University "Higher
 School of Economics", 101000 Moscow, Russia }
\affiliation{Institut f{\"u}r Nanotechnologie, Karlsruhe Institute of Technology, 76021 Karlsruhe, Germany}
\author{M. V. Feigel'man}
\affiliation{ L. D. Landau Institute for Theoretical Physics, Kosygin str.2, Moscow 119334, Russia}
\affiliation{Skolkovo Institute of Science and Technology, 143026 Skolkovo, Russia}
\affiliation{ Moscow Institute of Physics and Technology, Moscow 141700, Russia}

\begin{abstract}
We study stability of the SYK$_4$ model with a large but finite number of fermions $N$ with respect 
to a perturbation, quadratic in fermionic operators.  We develop  analytic perturbation theory in the 
amplitude of the SYK$_2$ perturbation and demonstrate  stability of the SYK$_4$
infra-red asymptotic behavior characterized by a Green function $G(\tau) \propto 1/\tau^{3/2} $,
with  respect to weak  perturbation. 
This result is supported by exact numerical diagornalization. 
Our results open the way to build a theory of non-Fermi-liquid states of strongly interacting fermions.
\end{abstract}
\maketitle

%%%%%%%%%%%%%%%%%%%%%%%%
%%%%%%%%%%%%%%%%%%%%%%%%
%%%%%%%%%%%%%%%%%%%%%%%%
%\section{Introduction}

The plenty of available data on various strongly correlated electronic materials~\cite{review1a,review1b}
calls for the development of a general theory of non-Fermi-liquid ground state(s) of an interacting many-body  fermionic
system. Still, no general theoretical scheme leading to such a behavior in the zero-temperature limit is known
(for a recent review see Ref.~\cite{review2}).
Mathematically, complexity of the problem is due to the absence of any general method to calculate 
non-Gaussian functional integrals which appear in the theory  of strongly interacting fermions. 

A new and fresh view on this old problem is provided by the recently proposed~\cite{sachdev2015bekenstein,Kitaev2015,Kitaev2017soft} 
Sachdev-Ye-Kitaev (SYK) model of interacting fermions. It has attracted a lot of attention recently as a possible boundary theory of a two-dimensional gravitational bulk \cite{Kitaev2015,polchinski2016spectrum,maldacena2016remarks}.
Original SYK model contains $N \gg 1 $ Majorana fermions, with the Hamiltonian consisting of a sum of all possible 4-fermion terms with random matrix elements $J_{ijkl} \sim J/N^{3/2}$ (note that the free (quadratic) term is missing in the SYK Hamiltonian). This model can be considered
as a non-linear generalization of  usual random-matrix Hamiltonians~\cite{guhr1998random}.  Furthermore, SYK$_q$
 models with arbitrary even $q=2k$ were introduced and studied~\cite{maldacena2016remarks}. These  models provide
 the most straightforward way to enhance the role of interaction between fermions, 
avoiding formation of any simply ordered structures which 
lead - usually, but not always~\cite{review2}  - to a breakdown of some evident
 symmetry of the Hamiltonian.

The SYK model is analytically tractable in the large-$N$ limit and shows two different types of asymptotic behavior for the fermionic Green function $G(\tau)$. In the intermediate time
range $1/J \ll \tau \ll t_c$, with $t_c \sim N/J$, the self-consistent approximation for interaction self-energy is valid and $G(\tau) \propto \tau^{-1/2}$ . For even larger times $\tau \gg t_c$, it was found in Ref.~\cite{bagrets2016sachdev} that fluctuations beyond the self-consistent treatment change the behavior of the Green function to $G(\tau) \propto \tau^{-3/2}$ (we treat exponentially large ergodic time-scale $\propto 2^{N/2}$  as being infinite).
Both these types of behavior are crucially different from the standard Fermi-liquid scaling
 $G(\tau) \propto 1/\tau$ which corresponds to nonzero  finite  density of low-energy states.
%the pole structure of the Green function in the energy representation. 
In other terms, low-energy excitations of the SYK model are not described by 
any kind of quasiparticles.

For the reasons described above, the SYK  model seems to be a very promising starting point to approach a theory of
non-Fermi-liquid ground state.  Few problems arise, however: i)  the absence of a quadratic term in the Hamiltonian
makes pure SYK Hamiltonian unrealistic for electronic systems; ii)  original SYK model contains Majorana fermions, which are quite scarce in Nature (see however few relevant proposals in Refs.\cite{PhysRevB.96.121119,pikulin2017black,chen2018quantum}); iii) most interesting properties of a non-Fermi-liquid state are those related to transport phenomena, while SYK is a random-matrix-type model without spatial coordinates.
Quite a number of recent publications address the issues listed above~\cite{Altman,DavisonSachdev,Yao,song2017strongly}.
Phase transitions controlled by the ratio of numbers of fermions in two SYK-like subsystems were studied in 
\cite{Altman,Ref1}. 

Generalization of the SYK model for complex fermions was developed in Refs.~\cite{DavisonSachdev,song2017strongly}.
A sequence of SYK "quantum dots" connected by weak (quadratic) tunnelling was considered in Refs.~\cite{Yao,song2017strongly},
making it possible to define and study transport quantities like resistance, thermal resistance, etc;
see also very recent extensive study in the same direction~\cite{Chowdhury2018}.  
Somewhat different direction was explored
in Refs.\cite{Ref2,Ref5}, where "dispersive SYK model" was introduced and studied in a way similar to Ref.\cite{song2017strongly}.
All these studies are restricted  by the use of self-consistent ($N\to \infty$) approximation, with the exception
provided by Ref.~\cite{Ref3}, where some renormalization group procedure based upon expansion over small
$q-2 \ll 1$ was formulated. Numerical analysis of the finite-range generalization of the SYK model with quadractic
terms was performed in Ref.~\cite{Ref4}, and analogies with Many-Body Localization were discussed.

However, all (known to us) studies of stability of SYK behavior w.r.t. to quadratic perturbations, indicate its runaway instability. As it was shown in Refs.~\cite{Altman,Yao,song2017strongly,Chowdhury2018} in the framework of the self-consistent approximation, the scaling dimension of the SYK$_2$ perturbation is negative when estimated within the conformal limit, corresponding to the time-scales $1/J \ll \tau \ll t_c$ . The papers~\cite{song2017strongly,Chowdhury2018} demonstrate an interesting non-Fermi-liquid behavior in the intermediate temperature region $T^* < T \ll J$, 
but still obtain Fermi-liquid behavior in the lowest $T$ range below $T^*$. 

In the present Letter we reconsider the  problem of the SYK$_4$ stability w.r.t. quadratic perturbations,
going beyond the saddle-point approximation. We  study fermionic Green function  in the region 
$\tau \gg t_c$ by means of perturbation theory in the amplitude of SYK$_2$ terms, using the infra-red asymptotic solution
~\cite{maldacena2016remarks,bagrets2016sachdev} as a starting point.
We show analytically that a weak SYK$_2$ perturbation does not change the $G(\tau) \propto 1/\tau^{3/2}$
asymptotics of the Green function,  but simply renormalizes the coefficient. 
This result proves the existence of a \textit{domain of stability}, with a non-zero area in the parameter
 space of Hamiltonians,  where a non-Fermi-liquid  is realized as a ground-state. We also perform numerical analysis of the Green function of the mixed SYK$_4$ + SYK$_2$ model to support our analytic study.
%%%%%%%%%%%%%%%%%%%%%%%%
%%%%%%%%%%%%%%%%%%%%%%%%
%%%%%%%%%%%%%%%%%%%%%%%%

%%%%%%%%%%%%%%%%%%%%%%%%
%%%%%%%%%%%%%%%%%%%%%%%%
%%%%%%%%%%%%%%%%%%%%%%%%
%\section
\textit{The model and basic equations.}
We consider the model defined by the following Hamiltonian
\begin{equation}
H=\frac{1}{4!}\sum_{i,j,k,l} J_{i,j,k,l}\chi_i\chi_j\chi_k\chi_l+\frac{i}{2!}\sum \Gamma_{i,j}\chi_i\chi_j
\label{Hamiltonian}
\end{equation}
where $\chi_i$ are Majorana fermions and all indices run from $1$ to $N$. The matrix elements $J_{ijkl}$ and  $\Gamma_{i,j}$ are fully antisymmetric and independent random Gaussian variables with zero mean and the variances $\langle J^2_{ijkl}\rangle=\frac{3!J^2}{N^3},$ $\langle\Gamma_{ij}^2\rangle=\frac{\Gamma^2}{N}$.
The functional integral representation of this theory is described by the
 action $S=-\frac{N}{2}\left(S_{1}+S_{2}\right)$ with two contributions\cite{maldacena2016remarks,bagrets2016sachdev}:
\begin{equation*}
S_{1}=\tr\log(\partial_\tau-\Sigma_{\tau\tau^\prime})+\int d\tau d\tau^\prime \left(\frac{J^2}{4}G^4_{\tau\tau^\prime}-\Sigma_{\tau\tau^\prime}G_{\tau^\prime\tau}\right),
\end{equation*}
and
\begin{equation}
\label{actionpert}
S_{2}=\int d\tau d\tau^\prime \frac{\Gamma^2}{2} G_{\tau\tau^\prime}^2.
\end{equation}
In the limit $N\gg1$ the mean-field analysis is appropriate and the corresponding saddle-point equations read
\begin{gather}
\label{saddle1}
\partial_\tau G_{\tau\tau^\prime}-\int d\tau^{\prime\prime}\Sigma_{\tau\tau^{\prime\prime}}G_{\tau^{\prime\prime}\tau^\prime}=\delta(\tau-\tau^\prime), \\ 
\Sigma_{\tau\tau^\prime}=J^2 G_{\tau\tau^\prime}^3+\Gamma^2 G_{\tau\tau^\prime}.
\label{saddle-point equations}
\end{gather}
We are going to study corrections to the SYK model Green function $G(\tau)$
assuming dimensionless parameter $\gamma =\Gamma/J$ to be small.
 Within applicability range of  the saddle-point Eqs. (\ref{saddle1}), (\ref{saddle-point equations}), the scaling dimension of the perturbation  is negative, $\Delta_\gamma = - \frac12$. As a result, $G_{SYK}^{(0)}(\tau) \propto (J\tau)^{-1/2}$ (the mean-field solution at $\gamma=0$) is unstable w.r.t. the perturbation: 
at $\tau \geq \tau^* \sim 1/J\gamma^2$ it is replaced by the usual Fermi-liquid behavior 
$G(\tau) \propto  (J\gamma\tau)^{-1} $.  On the other hand, at sufficiently long times $ t \gg t_c$,
soft re-parametrization modes~\cite{Kitaev2015,Kitaev2017soft,maldacena2016remarks} become relevant and the Green function of the pure SYK$_4$ 
model acquires different scaling~\cite{bagrets2016sachdev}

\begin{equation}
G(\tau)= \frac1{(4\pi)^{1/4}}\frac{1}{\sqrt{Jt_c}}\left(\frac{t_c}{\tau}\right)^{\frac{3}{2}} \equiv
\frac{\Gamma^4(\frac{1}{4})}{\sqrt{2MJ}\pi^{5/4}}\left(\frac{M}{\tau}\right)^{\frac{3}{2}}
\label{eq:syk0}
\end{equation}
For the reasons that become clear soon, we introduced  new notation 
$M = \pi t_c/\Gamma^{4}\left(\frac14\right)$, where $\Gamma(x)$ is the Euler gamma-function.

For sufficiently weak perturbation $\gamma \ll 1/\sqrt{N}$, the crossover timescale $\tau^*$ becomes larger than $t_c$
and loses its relevance: the analysis of  the SYK solution stability should now be performed using 
the asymptotic behavior (\ref{eq:syk0}) as a starting point.  Before we develop this analysis, a brief reminder
 on the origin of the result (\ref{eq:syk0}) is in order.
%%%%%%%%%%%%%%%%%%%%%%%%
%%%%%%%%%%%%%%%%%%%%%%%%

The saddle-point solution of  Eqs. (\ref{saddle1},\ref{saddle-point equations}) at $\Gamma=0$ is invariant
w.r.t. reparametrization of time, $\tau \to f(\tau)$, which is an approximate symmetry of the full action $S_1$,
see~\cite{sachdev2015bekenstein,Kitaev2015,Kitaev2017soft}.  Fluctuations around the saddle-point can be
accounted for by a kind of "sigma-model" defined on the the manifold of  functions $\phi(\tau)$,
defined via relation $df/dt = e^{\phi(t)}$.
This field theory has a very simple action\cite{bagrets2016sachdev}:
\begin{equation}
 S_\phi=\frac{M}{2}\int (\phi^\prime)^2 d\tau.
\label{action-soft}
\end{equation}
 Asymptotic behavior (\ref{eq:syk0}) of the Fermionic Green function 
$G_{\tau\tau^{\prime}}=\frac{1}{N}\sum_i\chi_i(\tau)\chi_i(\tau^{\prime})$ can then be obtained by the averaging of the functional
\begin{equation}
G_{\tau\tau^\prime}\left[\phi(\tau)\right] = 
\frac{1}{\sqrt{2J\sqrt{\pi}}}\sgn(\tau-\tau^\prime)\frac{e^{\phi(\tau)/4 }e^{\phi(\tau^\prime)/4 }}
{|\int\limits_{\tau^\prime}^{\tau}e^\phi(\tilde{\tau})d\tilde{\tau}|^{1/2}},
\label{correction_to_Green_function}
\end{equation}
 with the action (\ref{action-soft}), applicable in the long time limit $\tau \gg t_c$.
For actual calculations of functional integrals like Eq.(\ref{correction_to_Green_function})
with the action (\ref{action-soft}), we follow Refs.~\cite{bagrets2016sachdev,bagrets2017power} where
a very useful reduction to the  Liouville quantum mechanics~\cite{zamolodchikov1996conformal,shelton1998effective,teschner2001liouville,nakayama2004liouville}
was employed (see also Refs. \cite{Refc1,Refc2,kitaev2018statistical}).

There are various results in the literature~\cite{maldacena2016remarks,bagrets2016sachdev} 
concerning determination of the important parameter $M = M(N,J)$ which enter the action (\ref{action-soft}).
We prefer to employ the relation between results for the full Density of States of the SYK$_4$ model obtained i) via asymptotic 
low-energy theory, expressed in terms of $M$: $\rho(\epsilon)\propto \sinh\left(2\pi\sqrt{2M\epsilon}\right)$ \cite{bagrets2017power}, and ii) by the method of generalized orthogonal polynomials~\cite{garcia2017analytical}: $\rho(\epsilon)\propto \sinh\left(\frac{2\pi\sqrt{2}\sqrt{\epsilon/\epsilon_N}}{\ln 1/\eta_N}\right)$, were $\eta_N=1-\frac{32}{N}+O(\frac{1}{N^2})$ and $\bar{\epsilon}_N= \frac{J N}{16\sqrt{2}}+O(1)$ (at finite $N$ the expressions are given in the Ref. \cite{garcia2017analytical}). Comparison of two approaches yields
\begin{equation}
\label{MDef}
M=\frac{m(N)}{32\sqrt{2}}\frac{N}{J}\, 
%\quad \rm{where} \,\, \lim_{N \to \infty} m(N) =1.
\end{equation}
where interpolating function $m(N)$ approaches 1  in the limit $N=\infty$.
Note that convergence of $m(N)$ upon increase of $N$ is very slow; in particular, $m(32) \approx 0.54$.
Note also numerical factor $\sim 0.02$ in the RHS of  Eq.(\ref{MDef}), which makes $M$ much smaller than $N/J$.
Fortunately, the actual time-scale which enters Green functions $G(\tau)$ is 
$t_c = M\Gamma^{4}(\frac14)/\pi \approx 55 M \approx 1.2 N m(N)/J$; it will be important below  for the comparion with numerical data at large but finite $N = 32$.

\textit{Perturbation theory.}
First-order correction to the Green function $G(\tau)$ due to the quadractic term $S_2$ in the action can be found (see Supplementary Material for more details) in a straightforward way as follows 
(notation $\langle .. \rangle_0$ means the average  over  $\phi$ field with the  action $S_{\phi}$, see Eq. (\ref{action-soft})):
\begin{equation}
\delta G(\tau) =-\langle G_{\tau,0}[\phi]S_{2}[\phi]\rangle_0 + \langle G_{\tau,0}\rangle_0\langle S_{2}\rangle_0
\label{eq:pert}
\end{equation}
Substituting here Eq. (\ref{actionpert}), we find that the first term of Eq. (\ref{eq:pert}) contains an average
(over $\phi(\tau)$ fluctuations) of the product of three functionals like (\ref{correction_to_Green_function}), 
with the time arguments $0,\tau$ and $\tau_1,\tau_2$, where further integration over $\tau_{1,2}$ is implied.
Functional integration over $\phi(\tau)$ with the action (\ref{action-soft}) should be performed
 separately in 6 different time regions with the following order of time arguments:
\begin{center}
	\begin{tabular}{|c|c|c|c|c|c|}
		\hline
	 1 & 2 & 3 &4 &5 &6 \\ 
	 \hline
	\small $ \tau_2,\tau_1,0,\tau$ & \small  $\tau_2,0,\tau_1,\tau$ &  \small $\tau_2,0,\tau ,\tau_1$ & \small $0,\tau_2,\tau_1,\tau$ & \small $0, \tau_2, \tau,\tau_1$ & \small $0,\tau,\tau_2,\tau_1$   \\
	 \hline
	\end{tabular}
\end{center}
Domains 1 \& 6 have trivial structure and their contributions are canceled completely by the second term in Eq. (\ref{eq:pert}). 
 Combining other  contributions  with the corresponding parts of the second term in Eq.(\ref{eq:pert}), we find
\begin{equation}
\label{eq:G}
\delta G(\tau)=\frac{N\sqrt{MJ}}{4 \pi^{5/4}}\gamma^2\left[\sum_{i=2}^{5} f_i(\frac{\tau}{2M})-f_Z(\frac{\tau}{2M})\right].
\end{equation}
where functions $f_i(x)$ \, (for $i=2,3,4,5$)  and $f_Z(x)$ are defined and calculated in the Supplementary Material.
 In total, in the long-time limit $\tau\gg M$ we have: 
 \begin{eqnarray}
 \delta G(\tau) =  c N \sqrt{MJ}\gamma^2(\tau/M)^{-\frac{3}{2}}
 \end{eqnarray}
with $c\approx 108$. Comparing with Eq. (\ref{eq:syk0}), we find $\delta G/G\approx  3.7 JMN\gamma^2$
As a result
\begin{equation}
\label{eqres}
\delta G/G\approx  0.081 m(N) N^2\gamma^2,
\end{equation}
recall that $m(N)$ is defined in the Eq. (\ref{MDef}).
%Let us note that for analysis of numerical data it will be important that convergence of $m(N)$ to its thermodynamic limit is quite slow, in particular, $m(32)\approx 0.54$.
%%%%%%%%%%%%%%%%%%%%%%%%
%%%%%%%%%%%%%%%%%%%%%%%%
Equation (\ref{eqres}) demonstrates that relevant parameter of the perturbation theory in the infrared limit is 
actually $\gamma N \equiv b $, and perturbation of the SYK$_2$ type only modifies the numerical prefactor in $G(\tau)$.

%\textit{The role of quadratic terms near the saddle-point.}
% we present, for completeness, results of the saddle-point treatement of the action $S_1 + S_2$, assuming that
%$\beta = \Gamma/J \sim O(1) $ while $N \to \infty$.

%%%%%%%%%%%%%%%%%%%%%%%%
%%%%%%%%%%%%%%%%%%%%%%%%
%%%%%%%%%%%%%%%%%%%%%%%%
\textit{Numerical data}

\begin{figure}
\centering
 \includegraphics[width=0.5\textwidth]{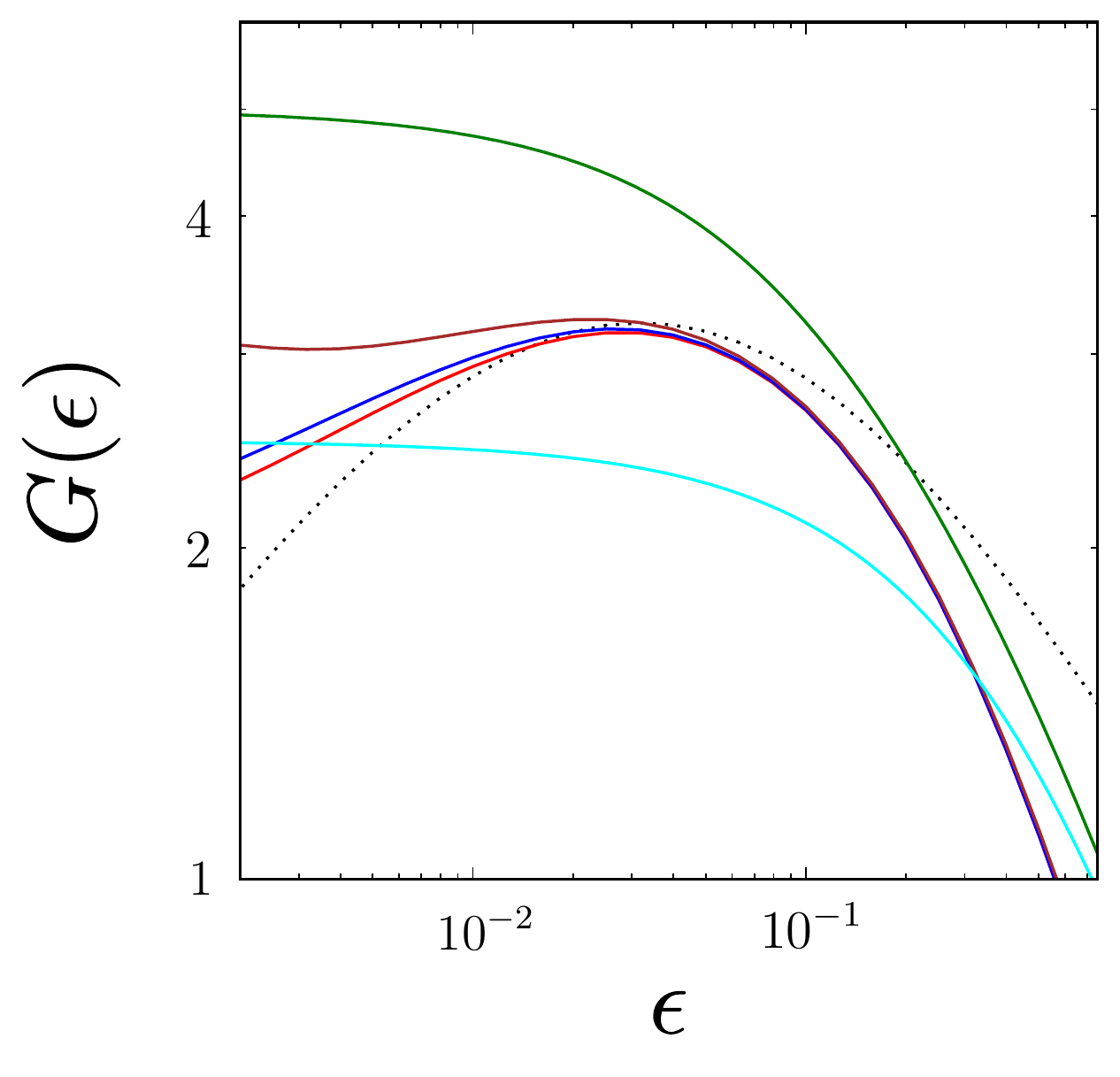}
\caption{Green function of the SYK model (log-log scale): i) red, blue, brown: exact diagonalization for $N=32$ fermions and $\gamma=0,0.01,0.03$;  ii) green, cyan: solution of mean-field equations Eqs. (\ref{saddle1}) and (\ref{saddle-point equations}) at $\gamma=0.2$ and $\gamma=0.4$; iii) dotted line: analytical result of Eq. (22) from Ref. \cite{bagrets2016sachdev}, interpolating between $\epsilon\gg M^{-1}$ and $\epsilon\ll M^{-1}$ limits, evaluated at $N=32$.}
\label{fig:g}
\end{figure}
Below we present numerical results for the Green function $G(\epsilon)$ of the SYK$_2$+SYK$_4$ model
 in the energy representation. This Green function  can be numerically studied with two complementary approaches: via exact diagonalization  at finite $N$ and directly in the limit of $N\to\infty$ via solution of the mean-field equations Eqs. (\ref{saddle1}) and (\ref{saddle-point equations}). Within numerical analysis below, we put $J=1$. For exact diagonalization, we consider the Hamiltonian of Eq.~(\ref{Hamiltonian}) for $N=32$ fermions for the range of $\gamma$ and average over hunderds of disorder realizations (we employ representation of Majorana algebra, used in Ref. \cite{haque2017eigenstate}). This results into red, blue and brown curves on the Fig. \ref{fig:g}. The most interesting regime is realized at small $\epsilon \leq 1/t_c$ (that is, below the maximum of the function $G(\epsilon)$). Characteristic time-scale $t_c \approx 20$ for $N=32$.  Unfortunately, the region  $\epsilon \leq 1/t_c$  is limited from below by the many-body energy scale 
$\epsilon =\epsilon_{MB}$ (defined as an energy, counted from the ground-state, where the many-body level spacing $\delta\epsilon(\epsilon)$ becomes comparable to $\epsilon$ itself.)
For $N=32$ the corresponding cut-off  $\epsilon_{MB} \approx 2\cdot 10^{-3}$ determines the left edge on the Fig.1.
The respective energy interval $\epsilon_{MB}\ll\eps\ll M^{-1}$ is not really large enough to admit for the predicted asymptotic behaviour $G(\eps)\propto\eps^{1/2}$: compare with the dotted line which is evaluated according to theoretical prediction of Eq. (22) of the Ref. \cite{bagrets2016sachdev}. However, qualitatively at $\gamma=0$ we find the expected behaviour. At small  $\gamma\ll N^{-1}$ the corresponding part of $G(\eps)$ dependence shifts up in the log-log
 plot (see blue curve for $\gamma = 0.01 \ll 1/N$) without change of  behavior as function of $\epsilon$, in agreement with  our analytical result (\ref{eqres}). However, at slightly larger values of $\gamma\sim 1/N$ (brown curve)
the Green function $G(\epsilon)$ saturates at low energies, in agreement with the Fermi-liquid behaviour. 

At even larger values of $\gamma \geq 1/\sqrt{N}$,  the asymptotic region with $G(\epsilon) \propto \sqrt{\epsilon} $  disappears completely, and Green function can be approximated by the solution of mean-field Eqs. (\ref{saddle1}) and (\ref{saddle-point equations}).  Interaction term is then important at higher energies $\epsilon \geq \epsilon_\gamma = 
\gamma^2 J > 1/t_c$ only.
The green and cyan curves show the numerical solutions to these equations. Note that mean-field solutions differ in the region $\epsilon \gg 1/t_c$  from the analytical result  given by Eq.(22) of Ref.~\cite{bagrets2016sachdev}  evaluated at $N=32$ (dotted line in Fig.\ref{fig:g}).  
These deviations appear since the asymptotic region $ 1/t_c = 0.05 \ll\eps\ll 1$ is apparently not wide enough. 
Thus, finite-size effects are detrimental for all analytically available asymptotics in this problem even for relatively large system of $N=32$ fermions. Although available system size is on border-line of emergence of respective asymptotic regions, we believe that the results of ED and mean-field studies are consistent with our analytical estimates. In particular, low-$\eps$ limits of $G(\eps)$ demonstrated
by cian and green lines are in agreement  with analytic result for pure SYK$_2$ theory,
$G (\eps \to 0) = 1/\Gamma $.

 %%%%%%%%%%%%%%%%%%%%%%%%
%%%%%%%%%%%%%%%%%%%%%%%%
%%%%%%%%%%%%%%%%%%%%%%%%

%%%%%%%%%%%%%%%%%%%%%%%%
%%%%%%%%%%%%%%%%%%%%%%%%
%%%%%%%%%%%%%%%%%%%%%%%%
\textit{Conclusions.}
Schematically, our results for the zero-temperature phase diagram of the combined SYK$_4$ - SYK$_2$ model
are shown in the Fig. \ref{fig:scheme}.  We emphasize somewhat unusual scaling limit of large $N$
that is employed here.  Namely, we consider $N \gg 1$ as some finite number, but we neglect  
exponentially small  many-body level spacing $\epsilon_{MB} \sim 2^{-N/2}$. 
Then our results demonstrate the presence of a \textit{phase transition}  between fully chaotic non-Fermi-liquid ground state
realized at $b \equiv \gamma N < b_c$, and  Fermi-liquid ground state existing at $b > b_c$, with $b_c \sim 1$.
We emphasize  that the corresponding critical value of the amplitude of the quadratic perturbation $\Gamma$ equals
 $\Gamma_c = b_c J/N$. In other terms, the effect of this perturbation in the infrared limit is much stronger 
than one could naively expect considering its effect at short times $t \leq t_c$ where relevant $\Gamma$ scales as $1/\sqrt{N}$. 

Note that $1/t^{3/2}$ long-time asymptotics of the Green function in pure SYK model
can be understood~\cite{Referee}  as a result of the square-root edge singularity of the full many-body DoS \cite{bagrets2017power,garcia2017analytical},  together with chaotic non-structured  nature of matrix elements that enter Lehman expansion for the Green function. Then, the phase transition we found upon increase of quadratic perturbation $b$ can be understood as a transition to non-chaotic state, with matrix elements aquiring nontrivial structure leading to Fermi-liquid type of behavior $G(t) \sim 1/t$.

\begin{figure}
\centering
 \includegraphics[width=0.5\textwidth]{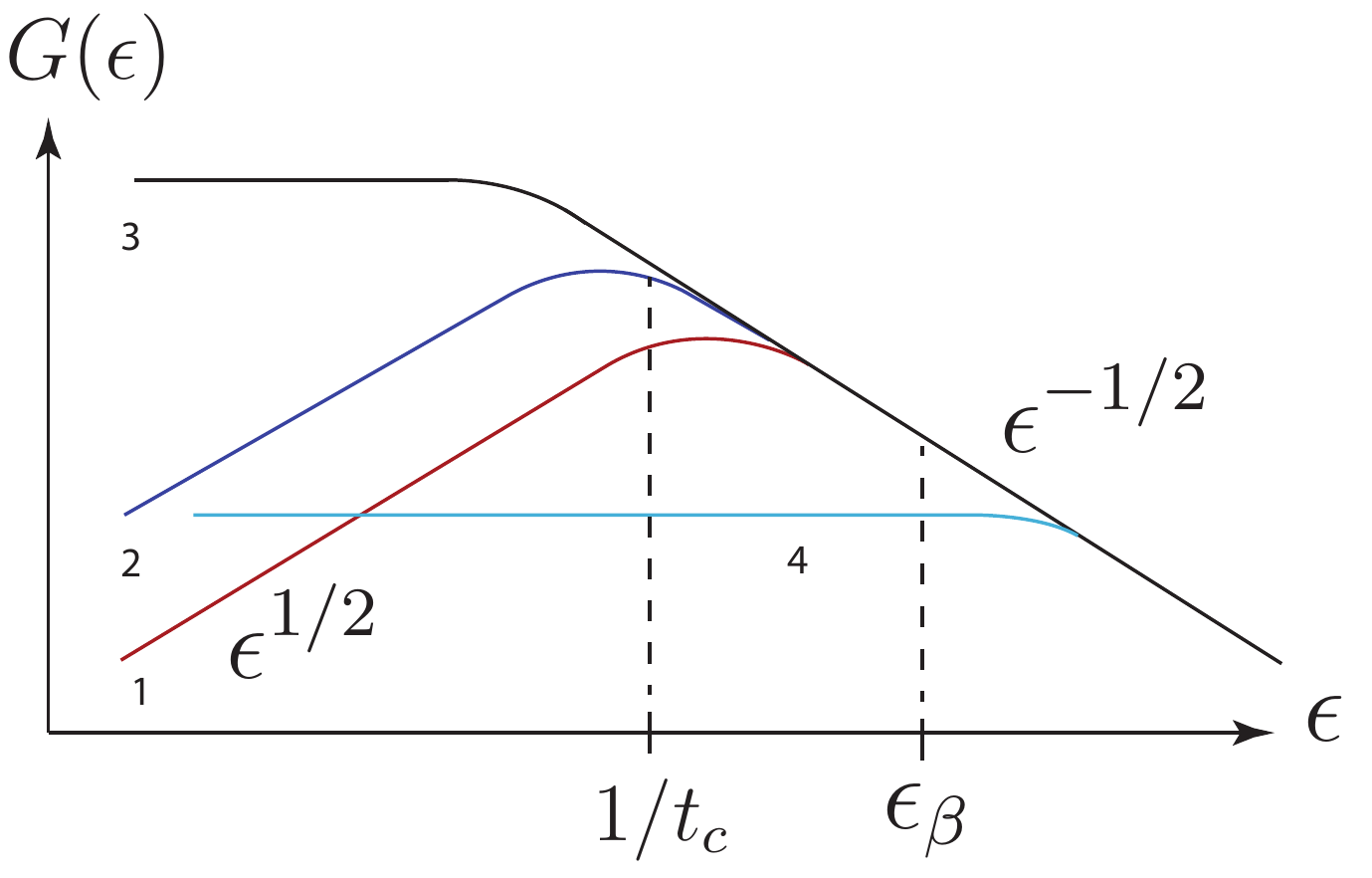}
\caption{Sketch of the Green function $G(\epsilon)$ in log-log scale, in the limit of $N\gg 1$ for 
several values of $b=\gamma N$ ordered as
$b_{1}<b_{2}<b_c<b_{3}<b_{4}$, with critical $b_c$ of the order unity.  NFL-FL transition occurs between
blue (2) and grey (3) lines. Light-blue line (4)  corresponds to large $b \geq \sqrt{N}$ when
4-fermion interaction is relevant at high energies above $ \epsilon_{\gamma} \gg 1/t_c$ only. }
\label{fig:scheme}
\end{figure}

Recently, chaotic-integrable transition for SYK model with quadratic perturbation was studied in Ref. \cite{garcia2018chaotic}. The authors have shown (judging it by Lyapunov exponent of the
 4-point out-of-time-order correlation function)  that for several values of  $\gamma \sim O(N^0)$,
 there exists a finite positive temperature $T=T_c(\gamma)$ such that at $T>T_c(\gamma)$ the system behaves 
chaotically, while at lower temperatures Lyapunov exponent drops to zero as it is expected 
for Fermi-liquid state with quasiparticle-based classification of eigenstates.
 We believe that this transition is of the same  kind as we found  at zero temperature for small $\gamma \sim b/N$. In this respect, see a recent study of energy-resolved spectral and many-body-wavefunction statistics, reported in Ref. \cite{nosaka2018thouless}. In particular,  Fig. 13a  demonstrates a qualitative change in the distribution function of the structural entropy of exact eigenstates already between $\kappa=0$ and $\kappa=1$  at low total energy, indicating the phase transition happening  at $\kappa\lesssim 1$ in agreement with our results.
% Indeed, we found that for small enough $\beta<\beta_c/N$, the system shows non-Fermi liquid behaviour down to zero temperature.

It would be very interesting to study similar Non-Fermi-Liquid - Fermi-Liquid  transition
 in a chain (or lattice)  of SYK-like "quantum dots". Note that for such extended models 
 there is no issue with finiteness of the non-zero many-body level spacing $\epsilon_{MB}$
 and  the problem of the NFL-FL phase transition can be formulated in the strict sense.
As was already mentioned in the Introduction,
 the transition of that kind was studied
in Refs.~\cite{song2017strongly,Chowdhury2018} at non-zero temperatures within a $N \to \infty$ limit.
Another approach  was developed  in Ref.~\cite{dai2018global} where an effect of the  SYK-like interaction
upon the properties of a random-hopping chain was investigated via  numerical analysis of the
level statistics. The authors of~\cite{dai2018global} found the MBL-type transition between fully localized
and ergodic ground states at rather \textit{low and decreasing with} $N$ ratio $J/\Gamma$ of the SYK coupling
to quadratic coupling;  thus it seems to be qualitatively different from the transition we found for 
a single SYK system. 

Finally, we would like to mention an interesting physical problem which may bear some resemblance with
SYK$_4$ - SYK$_2$ model considered here.  It is well known that strong  potential  disorder 
suppresses superconducting  transition with non-trivial (d-wave or p-wave) pairing, due to 
random mixing of electron states between different sectors of the Fermi-surface.  
However, mean-square magnitude of the (random-sign) Cooper interaction amplitude survives impurity
scattering. It means that electron states with energies exactly at the Fermi-surface  constitute
a kind of SYK-type model with a random 4-fermion interaction.  Potentially, this interaction may occur
to be strong enough to lead to a non-Fermi-liquid ground state without formation of any order parameter.

%\com{Possible physical implementations of the SYK model\cite{PhysRevB.96.121119,pikulin2017black,chen2018quantum} should be mentioned.}
%\textbf{MF:  I am not sure those refs. are so much needed for us; they are all very much speculative and
%difficult to grasp.}

%%%%%%%%%%%%%%%%%%%%%%%%
%%%%%%%%%%%%%%%%%%%%%%%%
%%%%%%%%%%%%%%%%%%%%%%%%

%%%%%%%%%%%%%%%%%%%%%%%%
%%%%%%%%%%%%%%%%%%%%%%%%
%%%%%%%%%%%%%%%%%%%%%%%%
%\section{Acknowledgments}
	 
We thank A. Yu. Kitaev, and D. A. Bagrets for many useful discussions. K. S. T. acknowledges support from the Foundation for Development of Theoretical Physics and Mathematics “Basis” and from Alexander von Humboldt Foundation. The work is partially supported by the Russian Academy of Sciences program "Modern problems of low-temperature physics".
%%%%%%%%%%%%%%%%%%%%%%%%
%%%%%%%%%%%%%%%%%%%%%%%%
%%%%%%%%%%%%%%%%%%%%%%%%

\bibliography{syk}
\newpage 
\begin{widetext}
	
	\begin{center}{\large \bf Supporting Material "SYK model with quadratic perturbations: the route to a non-Fermi-liquid."}
	\end{center}
	\maketitle
	\section{Plan of calculation}
	We will present method for evaluation of various correlation function for general SYK$_q$ model with $q$ simultenously
	interacting fermions; then we  will  put $q=4$ in the end of the calculations. 
	The low-energy limit of  SYK$_q$ model is described by the  "sigma-model" action  over the manifold of monotonic functions 
	$f(\tau)$, which corresponds to  the  re-parametrization symmetry  of the mean-field solution in the scaling limit.
	The functions defined on this manifold can be conveniently parametrised in terms of the field $\phi(\tau)$,
	which is defined according to $f^\prime(\tau)=e^{\phi(\tau)}$. In such a representation,  the action  reduces 
	to the simple form
	\be
	S=-\frac{M}{2}\int (\phi^\prime)^2 d\tau
	\label{actionM}
	\ee 
	where parameter $M$ depends on the number of fermions $N$ and on the value of $q$. In such a parametrization the measure of the functional integration is flat. Note that here and below in the SM we put interaction strength $J=1$.
	The field $G(\tau,\tau')$, which becomes equal to the fermionic Green function 
	upon integrating over $\phi(\tau)$, reads in the parametrization as follows:
	\begin{eqnarray}
	G(\tau,\tau^\prime)=sign(\tau-\tau^\prime)b^\Delta \frac{e^{\Delta\phi(\tau)}e^{\Delta\phi(\tau^\prime)}}
	{|\int_{\tau^\prime}^{\tau}e^{\phi(\tilde{\tau})}d\tilde{\tau}|^{2\Delta}} 
	\label{G2}
	\end{eqnarray}
	with $\Delta=\frac{1}{q}$ and $b=(\frac{1}{2}-\Delta)\frac{\tan(\pi\Delta)}{\pi}$.
	To simplify the notation in the following,  we will consider averaging of the objects like  $G^{\frac{n}{\Delta}}$.  
	The average Green function will come up as a specific result at  $n=\Delta$.
	
	Below in Sec. II we  rederive some results from Ref.[\onlinecite{bagrets2016sachdev}] in a slightly different way;
	namely, we show how to reduce evaluation of the Green function (\ref{G2}) with the action (\ref{actionM})
	to the calculation of matrix elements of the Liouville quantum mechanics.
	
	In  Sec. III we evaluate correlation functions of the products of various powers of Green functions
	$G^\frac{n}{\Delta}$ which are necessary to calculate the corrections to the Green function generated by 
	the perturbation of the SYK$_2$ type:
	% can be reduced to evaluation of corrections ot the averages of $G^\frac{n}{\Delta}$ 
	\begin{eqnarray}
	\label{corr}
	\delta\langle G^{n/\Delta}_{\tau_1,\tau_2}\rangle=-\langle G^{n/\Delta}_{\tau_1,\tau_2} S_{int}\rangle+\langle G^{n/\Delta}_{\tau_1,\tau_2} \rangle \langle S_{int}\rangle.
	\end{eqnarray}
	\be
	S_{int}=-\frac{\Delta N\Gamma^2}{2 m}\int d\tau_1 d\tau_2 G^{m/\Delta}(\tau_1,\tau_2)
	\label{Sint}
	\ee. 
	Finally we will be interested in the case $m=\frac{1}{2}$ and $n=\frac{1}{4}$, $\Delta=\frac{1}{4}$. 
	
	To simplify further formulae, we switch to dimensionless time units $t=\frac{\tau}{2M}$ and
	introduce new notation for the Green function:
	\be
	g_n(t,t^\prime)= \left(\frac{1}{\Gamma(2 n)}\frac{b^{ n}}{(2M)^{2 n}}\right)^{-1} G^{n/\Delta}(\tau,\tau^\prime)
	\ee
	Therefore Eq.(\ref{corr}) with $S_{int}$ from Eq.(\ref{Sint}) can now be rewritten in the form
	\begin{eqnarray}
	\delta\langle G^{n/\Delta}_{\tau_1,\tau_2}\rangle=
	\frac{N \Gamma^2\Delta}{m} \frac{b^{(n+m)}}{\Gamma(2 n)\Gamma(2 m)(2M)^{2 (n+m)-2}}\int_{t_3>t_4} dt_3 dt_4 \left[\langle g_n(t_1,t_2)g_m(t_3,t_4)\rangle-\langle g_n(t_1,t_2)\rangle \langle g_m(t_3,t_4)\rangle\right]
	\label{Gnd}
	\end{eqnarray}

	%%%%
	%%%%For comparison we need:
	%%%%\begin{eqnarray}
	%%%%\frac{\delta \langle G^n(\tau,\tau^\prime)\rangle}{\langle G^n(\tau,\tau^\prime)\rangle}=
	%%%%\frac{N \Gamma^2}{m} \frac{b^{\Delta m}}{\Gamma(2\Delta m)(2M)^{2\Delta m-2}}
	%%%%%%%%\frac{1}{\langle g_n(t,t^\prime) \rangle}\int_{t_3>t_4} dt_3 dt_4 \left[\langle g_n(t_1,t_2)g_m(t_3,t_4)\rangle-%%%%\langle g_n(t_1,t_2)\rangle \langle g_m(t_3,t_4)\rangle\right]
	%%%%\end{eqnarray}
	
	To evaluate expression (\ref{Gnd}) one has to calculate the average 
	$\langle g_n(t_1,t_2) g_m ({t_3,t_4})\rangle$. The  difficulty of this calculation  is due to the absence of Wick contraction rules, so  all possible time orderings have to be considered explicitely. Symmetry of Green function  allows  us
	to fix relations $t_1>t_2$ and $t_3>t_4$, leaving 6 possible  time orderings: 1) $t_1>t_2>t_3>t_4$, 2) $t_1>t_3>t_2>t_4$, 3)$t_3>t_1>t_2>t_4$, 4) $t_1>t_3>t_4>t_2$, 5) $t_3>t_1>t_4>t_2$ and 6) $t_3>t_4>t_1>t_2$. 
	The orderings 1 and 6 are trivial \cite{bagrets2016sachdev}:
	\be
	\langle g_n(t_1,t_2) g_m(t_3,t_4)\rangle = \langle g_n(t_1,t_2)\rangle \langle g_m(t_3,t_4)\rangle
	\ee
	In  Sec. III we present  evaluation of the average values corresponding to the remaining four variants of time ordering. 
	In the  remaining Secs. IV - VI  we combine various contributions in the long-time limit and derive the final result.

	\section{Averaging various powers of the Green function}
	
	First of all, we rederive some results from Ref. \onlinecite{bagrets2016sachdev} in a slightly different way.
	We start from the formula for average power of the Green function
	\begin{eqnarray}
	\langle G^{n/\Delta}(\tau,\tau^\prime)\rangle=\int D\phi~b^{n\Delta} \frac{e^{n \phi(\tau)}e^{n \phi(\tau^\prime)}}{|\int_{\tau^\prime}^{\tau} e^{\phi(\tilde{\tau})}d\tilde{\tau}|^{2n}} e^{-\frac{M}{2}\int (\phi^\prime)^2d\tau }
	\end{eqnarray}
	Switching to dimensionless time, we write
	\begin{eqnarray}
	\label{gdef}
	\langle G^{n/\Delta}(t,t^\prime)\rangle=\int D\phi~\frac{b^{n}}{(2M)^{2n}} \frac{e^{n \phi(t)}e^{n \phi(t^\prime)}}{|\int_{t^\prime}^{t} e^{\phi(\tilde{t})}d\tilde{t}|^{2n}} e^{-\frac{1}{4}\int (\phi^\prime)^2 dt} 
	\end{eqnarray}
	Using identity  $\frac{1}{a^{2n}}=\int_0^\infty \frac{\alpha^{n-1}}{\Gamma(2n)} e^{-\alpha a}d\alpha$
	one can rewrite above expression as follows:
	\begin{eqnarray}
	\langle G^n(t,t^\prime)\rangle=\int_0^\infty \frac{\alpha^{2n-1}d\alpha}{\Gamma(2n)}\int D\phi~\frac{b^{n}}{(2M)^{2n}}e^{n \phi(t)}e^{n \phi(t^\prime)} e^{-\frac{1}{4}\int (\phi^\prime)^2 dt-\alpha \int_{t^\prime}^{t} e^{\phi(\tilde{t})}d\tilde{t}} 
	\end{eqnarray} 
	
	Functional integral over $\phi(t) $ can be interpreted as a quantum mechanical amplitude and evaluated explicitly. 
	There is a technical problem however: the field $\phi(t)$ in Eq. (\ref{gdef}) can be shifted by a constant: $\phi(t)\rightarrow \phi(t) +\phi_0$,  producing a divergent integral. In the calculation provided in Ref. \onlinecite{bagrets2016sachdev}, this zero mode appeared as an infinite multiplicative constant, coming from divergent integration
	over parameter $\alpha$.
	This divergence was argued~\cite{bagrets2016sachdev} to be irrelevant since it is related to
	the symmetry of the action. Slightly different formulation of the same approach is to put $\alpha$ equal to
	1 instead of integration over $\alpha$.
	% {\color{blue} Authors told that these integrals should be ignored  because this divergence connected with the symmetry of the action. This ignoring can be done in the following way: we should put $\alpha$ equal to 1 instead of integration over $\alpha$.} 
	Here we first check this idea by using another method. Namely,  we fix the "gauge condition" by putting the value 
	of $\phi(t)$  equal to $\phi_0$ and then integrate over  $\alpha$;  we obtain then the same result as in Ref.\onlinecite{bagrets2016sachdev}.   Therefore in our further calculations we will follow the approach of 
	Ref.~\cite{bagrets2016sachdev} which is simpler in implementation.
	
	Rewriting formulae  in terms of $g_n(t,t^\prime)$, we find
	\begin{eqnarray}
	\langle g_n(t,t^\prime)\rangle=\int_0^\infty \alpha^{2n-1}  d\alpha \int d\phi_1 \langle\phi_0|e^{n\phi}U_\alpha(t,t^\prime)e^{n\phi}|\phi_1\rangle
	\label{QM_everage_for_g_n},
	\end{eqnarray}
	where $U_\alpha(\tau,\tau^\prime)$  is the evolution operator corresponding to the Liouville's Hamiltonian $H=-\frac{1}{4}\partial^2_\phi+\alpha e^\phi$. It can be written as
	\begin{eqnarray}
	U_\alpha(t,t^\prime)=\int_0^\infty \frac{dk}{2\pi} e^{-k^2 (t-t^\prime)} |k,\alpha\rangle\langle k,\alpha|
	\end{eqnarray}
	with eigenstates
	\be
	\langle \phi|k,\alpha\rangle = \frac{2}{\Gamma(2ik)} K_{2ik}(2\sqrt{\alpha e^\phi}).
	\ee 
	It is more convenient to work with Mellin transformed eigenfunctions
	\begin{eqnarray}
	\frac{2}{\Gamma(2ik)}K_{2ik}(2\sqrt{x})=\int_{c-i\infty}^{c+i\infty } \frac{\Gamma(p-ik)\Gamma(p+ik)}{\Gamma(2ik)} x^{-p} \frac{dp }{2\pi i}.
	\end{eqnarray}
	We now introduce the "matrix element" $G(p,k)$ as follows:
	\begin{eqnarray}
	G(p,k)=\frac{\Gamma(p-ik)\Gamma(p+ik)}{\Gamma(2ik)}.
	\end{eqnarray}
	Then Eq. (\ref{QM_everage_for_g_n}) becomes
	\begin{eqnarray}
	\langle g_n(t,t^\prime)\rangle=\int_0^\infty \frac{dk}{2\pi}  e^{-k^2 (t-t^\prime)} \int_0^\infty \alpha^{2n-1}  d\alpha \int_{-\infty}^\infty d\phi_1 e^{n\phi_0}e^{n\phi_1}\langle\phi_0|k,\alpha\rangle\langle k,\alpha|\phi_1\rangle = \nonumber \\ =\int_0^\infty \frac{dk}{2\pi}  e^{-k^2 (t-t^\prime)} \int_0^\infty \alpha^{2n-1}  d\alpha \int_{-\infty}^\infty d\phi_1 e^{n\phi_0}e^{n\phi_1} \int \frac{dp_1}{2\pi i} \frac{dp_1}{2\pi i} G(p_1,k) G(p_1,-k) \alpha^{-p_1}e^{-p_1\phi_0} \alpha^{-p_2}e^{-p_2\phi_1}= \\ \nonumber = \int_0^\infty \frac{dk}{2\pi}  e^{-k^2 (t-t^\prime)} G(n,k) G(n,-k) 
	\end{eqnarray}
	In the limit of $t\gg 1$ we find
	\be
	\langle g_n(t,t^\prime)\rangle=\int_0^\infty \frac{dk}{2\pi}e^{-k^2(t-t^\prime)}\Gamma^4(\Delta n)(4k^2)=
	\frac{\Gamma^4(n)}{2\sqrt{\pi}(t-t^\prime)^\frac{3}{2}}.
	\ee
	which coinsides with the result of Ref.~\onlinecite{bagrets2016sachdev}.
	
	\section{Averaging the products of various powers of Green function.}
	
	We now turn to the calculation of the averages of the type $\langle g^n(t_1,t_2) g^m(t_3,t_4)\rangle$. Following the same steps as in the Sec. II we obtain
	\begin{eqnarray}
	\langle g_n(t_1,t_2) g_m(t_3,t_4)\rangle= \int_0^\infty \alpha^{2n-1} d\alpha\int_0^\infty \beta^{2m-1}d\beta\int D\phi e^{n\phi_1}e^{n\phi_2}e^{m\phi_3}e^{m\phi_4}e^{-\frac{1}{4}\int (\phi^\prime)^2dt-\alpha \int_{t_2}^{t_1} e^{\phi(\tilde{t})}d\tilde{t}-\beta \int_{t_4}^{t_3} e^{\phi(\tilde{t})}d\tilde{t}},
	\end{eqnarray}
	with shorthand notation $\phi_i=\phi(t_i)$. Like  in  Sec. II, we interprete the functional integral over $\phi(t)$ as a quantum-mechanical amplitude.  It is convenient to fix the "gauge" by setting $\alpha\to 1$, to simplify calculations. 
	The result of averaging  depends crucially on the specific  time ordering (see discussion in  Sec. I). 
	We present here  details of the calculation for the cases 2 and 3.  Results for the cases 4 and 5 can be obtained in  similar way, so we will provide the results only. 
	
	\subsection{Time-ordering 2: $t_1>t_3>t_2>t_4$}
	
	Quantum mechanical representation of the problem corresponds to the free particle motion at times $t < t_4$.  
	In the range of times  $t_4 < t < t_2$ the exponential potential $e^\phi$ with the magnitude equal to  $\beta$ is turned on, so the evolution during this time interval  is described by $U_\beta(t_2,t_4)$ (see Sec. II for the definition of $U(t,t')$). Next, in the time region between  $t_2$ and $t_3$, the evolution is governed by $U_{1+\beta}(t_3,t_2)$,
	and  between $t_1$ and $t_3$ it is given by $U_1(t_1,t_3)$. Finally, at  $t> t_1$ the particle is free again.
	
	Than quantum-mechanical average is of  the following  form 
	(hereafter we use  Roman subscripts to denote specific  time ordering,  which is the  2nd one currently):
	\begin{eqnarray}
	\langle g_n(t_1,t_2) g_m(t_3,t_4)\rangle_{II}= \int_0^\infty \beta^{2m-1}d\beta \int d\phi_1 d\phi_2 d\phi_3 d\phi_4 e^{n\phi_1}e^{n\phi_2}e^{m\phi_3}e^{m\phi_4} \nonumber \\ 
	\langle \phi_1|U_1(t_1,t_3)|\phi_3\rangle \langle\phi_3|U_{1+\beta}(t_3,t_2)|\phi_2\rangle\langle \phi_2|U_{\beta}(t_2,t_4)|\phi_4\rangle
	\end{eqnarray}
	Using explicit representation for $U$ we find
	\begin{eqnarray}
	\langle g_n(t_1,t_2) g_m(t_3,t_4)\rangle_{II}= \int_0^\infty \frac{dk_1 dk_2dk_3}{(2\pi)^3} e^{-k_1^2t_{1,3}}e^{-k_2^2t_{3,2}}e^{-k_3^2t_{2,4}}\int_0^\infty \beta^{2m-1}d\beta \nonumber \\ 
	\int d\phi_1 d\phi_2 d\phi_3 d\phi_4 e^{n\phi_1}e^{n\phi_2}e^{m\phi_3}e^{m\phi_4}\langle \phi_1|k_1,1\rangle \langle k_1,1|\phi_3\rangle \langle\phi_3|k_2,1+\beta\rangle \langle k_2,1+\beta|\phi_2\rangle\langle \phi_2|k_3, \beta\rangle \langle k_3,\beta|\phi_4\rangle
	\end{eqnarray}
	The integrations over $\phi$ are factorized. Integrations over $\phi_1$ and $\phi_4$ are trivial:
	\begin{eqnarray}
	\int d\phi_1 e^{n\phi_1} \langle \phi_1|k_1,1\rangle = \int d\phi_1 e^{n\phi_1} \int_{c-i\infty}^{c+i\infty} G(p,k_1)e^{-p\phi_1}\frac{dp}{2\pi i}=G(n,k_1)
	\end{eqnarray}  
	\begin{eqnarray}
	\int d\phi_4 e^{m\phi_4} \langle k_3,\beta|\phi_4\rangle = \int d\phi_4 e^{m\phi_4} \int_{c-i\infty}^{c+i\infty}  G(p,-k_3) e^{-\phi_4 p} \beta^{-p} \frac{dp}{2\pi i}= G(m,-k_3) \beta^{-m}
	\end{eqnarray}
	Integrations over $\phi_3$ and $\phi_2$ are more involved. Integrating over $\phi_3$ we find 
	\begin{eqnarray}
	\int d\phi_3 e^{m\phi_3}\beta^{m} \langle k_1,1|\phi_3\rangle \langle\phi_3|k_2,1+\beta\rangle=\int d\phi_3 e^{m\phi_3}\beta^m \int_{c-i\infty}^{c+i\infty} G(p_1,-k_1)G(p_2,k_2)e^{-\phi_3 (p_2+p_1)}(1+\beta)^{-p_2} \frac{dp_1 dp_2}{(2\pi i)^2 }= \nonumber \\  \beta^m\int_{c-i\infty}^{c+i\infty} G(m-p,-k_1)G(p,k_2)(1+\beta)^{-p} \frac{ dp}{2\pi i }= \nonumber \\(1+\beta)^{ik_2}\beta^m \frac{\Gamma(2ik_1)}{\Gamma(2ik_2)} G(m+ik_2,k_1)G(m-ik_2,-k_1)\bm{F}(m+ik_2-ik_1,m+ik_2+ik_1,2m,-\beta)= \nonumber \\ 
	(1+\beta)^{ik_2}\frac{\Gamma(2ik_1)}{\Gamma(2ik_2)}G(m-ik_2,-k_1)\int_{c_m-i\infty}^{c_m+i\infty}G(p+ik_2,k_1)\frac{\Gamma(m-p)}{\Gamma(m+p)} \beta^p \frac{dp}{2\pi i} \nonumber \\
	\end{eqnarray}
	with $\bm{F}(a,b,c,z)$ for the normalized Hyperheometric function, $c>0$ and $c_m \in (0,m)$. Integration over $\phi_2$ can be performed in the similar manner. We note a useful identity:
	\begin{eqnarray}
	\int d\phi \beta^m e^{m\phi} \langle k_L,\alpha|\phi\rangle \langle\phi|k_R,\alpha+\beta\rangle =\int d\phi \beta^m e^{m\phi} \langle -k_R,\alpha+\beta|\phi\rangle \langle\phi|-k_L,\alpha\rangle \nonumber \\ (1+\frac{\beta}{\alpha})^{ik_R}\frac{\Gamma(2ik_L)}{\Gamma(2ik_R)}G(m-ik_R,-k_L)\int_{c_m-i\infty}^{c_m+i\infty}G(p+ik_R,k_L)\frac{\Gamma(m-p)}{\Gamma(m+p)}\left(\frac{\beta}{\alpha}\right)^p \frac{dp}{2\pi i}
	\label{triangle}.
	\end{eqnarray}
	With this identity, we find
	\begin{eqnarray}
	\langle g_n(t_1,t_2) g_m(t_3,t_4)\rangle_{II}= \int_0^\infty \frac{dk_1 dk_2dk_3}{(2\pi)^3} e^{-k_1^2t_{1,3}}e^{-k_2^2t_{3,2}}e^{-k_3^2t_{2,4}}\int_0^\infty \beta^{-1} G(n,k_1)G(m,-k_3)d\beta  
	\nonumber \\ 
	(1+\beta)^{ik_2}\frac{\Gamma(2ik_1)}{\Gamma(2ik_2)}G(m-ik_2,-k_1)\int_{c_m-i\infty}^{c_m+i\infty}G(p+ik_2,k_1)\frac{\Gamma(m-p)}{\Gamma(m+p)}\beta^p \frac{dp}{2\pi i} \nonumber \\
	\beta^{ik_2}(1+\beta)^{-ik_2}\frac{\Gamma(-2ik_3)}{\Gamma(-2ik_2)}G(n+ik_2,k_3)\int_{c_n-i\infty}^{c_n+i\infty}G(q-ik_2,-k_3)\frac{\Gamma(n-q)}{\Gamma(n+q)}\beta^{-q} \frac{dq}{2\pi i}.
	\end{eqnarray}
	The last integration over $\beta$ gives:
	\begin{eqnarray}
	\langle g_n(t_1,t_2) g_m(t_3,t_4)\rangle_{II}= \int_0^\infty \frac{dk_1 dk_2dk_3}{(2\pi)^3} e^{-k_1^2t_{1,3}}e^{-k_2^2t_{3,2}}e^{-k_3^2t_{2,4}}  
	\nonumber \\ 
	G(n,k_1)G(m,-k_3)\frac{\Gamma(2ik_1)}{\Gamma(2ik_2)}\frac{\Gamma(-2ik_3)}{\Gamma(-2ik_2)}G(n+ik_2,k_3)G(m-ik_2,-k_1) \nonumber \\
	\int_{c_{\min(m,n)}-i\infty}^{c_{\min(m,n)}+i\infty}G(p+ik_2,k_1)\frac{\Gamma(m-p)}{\Gamma(m+p)}G(p,-k_3)\frac{\Gamma(n-p-ik_2)}{\Gamma(n+p+ik_2)}\frac{dp}{2\pi i} 
	\end{eqnarray}
	\subsection{Time-ordering 3:  $t_3>t_1>t_2>t_4$}
	Following the same steps as above we come to
	\begin{eqnarray}
	\langle g_n(t_1,t_2) g_m(t_3,t_4)\rangle_{III}= \int_0^\infty \beta^{2m-1}d\beta \int d\phi_1 d\phi_2 d\phi_3 d\phi_4 e^{n\phi_1}e^{n\phi_2}e^{m\phi_3}e^{m\phi_4} \nonumber \\ 
	\langle \phi_3|U_\beta(t_3,t_1)|\phi_1\rangle \langle\phi_1|U_{1+\beta}(t_1,t_2)|\phi_2\rangle\langle \phi_2|U_{\beta}(t_2,t_4)|\phi_4\rangle.
	\end{eqnarray}
	With expicit expression for $U$, we find
	\begin{eqnarray}
	\langle g_n(t_1,t_2) g_m(t_3,t_4)\rangle_{III}= \int_{0}^{\infty}\frac{dk_1 dk_2 dk_3}{(2\pi)^3}e^{-k_1^2 t_{3,1}}e^{-k_2^2 t_{1,2}}e^{-k_3^2 t_{2,4}}\int_0^\infty \beta^{2m-1}d\beta \nonumber \\ 
	\int d\phi_1 d\phi_2 d\phi_3 d\phi_4 e^{n\phi_1}e^{n\phi_2}e^{m\phi_3}e^{m\phi_4} \langle \phi_3|k_1,\beta \rangle \langle k_1,\beta|\phi_1\rangle \langle\phi_1|k_2,1+\beta \rangle \langle k_2,1+\beta|\phi_2\rangle\langle \phi_2|k_3,\beta \rangle \langle k_3,\beta|\phi_4\rangle.
	\end{eqnarray}
	Using the idenitty in Eq. (\ref{triangle}) we integrate over $\phi$:
	\begin{eqnarray}
	\langle g_n(t_1,t_2) g_m(t_3,t_4)\rangle_{III}= \int_{0}^{\infty}\frac{dk_1 dk_2 dk_3}{(2\pi)^3}e^{-k_1^2 t_{3,1}}e^{-k_2^2 t_{1,2}}e^{-k_3^2 t_{2,4}}\int_0^\infty \beta^{-1}d\beta  G(m,k_1)G(m,-k_3)\nonumber \\ 
	(1+\frac{1}{\beta})^{-ik_2}\frac{\Gamma(-2ik_3)}{\Gamma(-2ik_2)}G(n+ik_2,k_3)\int_{c_n-i\infty}^{c_n+i\infty}G(q-ik_2,-k_3)\frac{\Gamma(n-q)}{\Gamma(n+q)}\left(\frac{1}{\beta}\right)^q \frac{dq}{2\pi i} \nonumber \\
	(1+\frac{1}{\beta})^{ik_2}\frac{\Gamma(2ik_1)}{\Gamma(2ik_2)}G(n-ik_2,-k_1)\int_{c_n-i\infty}^{c_n+i\infty}G(p+ik_2,k_1)\frac{\Gamma(n-p)}{\Gamma(n+p)}\left(\frac{1}{\beta}\right)^p \frac{dp}{2\pi i}.
	\end{eqnarray}
	Finally, $\beta$-integration gives
	\begin{eqnarray}
	\langle g_n(t_1,t_2) g_m(t_3,t_4)\rangle_{III}= \int_{0}^{\infty}\frac{dk_1 dk_2 dk_3}{(2\pi)^3}e^{-k_1^2 t_{3,1}}e^{-k_2^2 t_{1,2}}e^{-k_3^2 t_{2,4}} G(m,k_1)G(m,-k_3)\nonumber \\ 
	\frac{\Gamma(-2ik_3)}{\Gamma(-2ik_2)}\frac{\Gamma(2ik_1)}{\Gamma(2ik_2)}G(n-ik_2,-k_1)G(n+ik_2,k_3)\int_{c_n-i\infty}^{c_n+i\infty}G(-p-ik_2,-k_3)
	G(p+ik_2,k_1)\frac{dp}{2\pi i}
	\end{eqnarray}
	
	In fact, for time-ordering 3 we can go even further and calculate one of the momentum integrals analytically: $p$-integration gives the momentum conservation law $2\pi \delta(k_1-k_3)$ and as a result:
	\begin{eqnarray}
	\langle g_n(t_1,t_2) g_m(t_3,t_4)\rangle_{III}= \int_{0}^{\infty}\frac{dk_1 dk_2 }{(2\pi)^2}e^{-k_1^2 t_{3,1}}e^{-k_2^2 t_{1,2}}e^{-k_1^2 t_{2,4}} G(m,k_1)G(m,-k_1) G(n-ik_1,-k_2)G(n+ik_1,k_2).
	\end{eqnarray}
	\subsection{Results for time-orderings 4 and 5}
	For the orderings 4 and 5 we provide only the results:
	\begin{eqnarray}
	\langle g_n(t_1,t_2) g_m(t_3,t_4)\rangle_{IV}= \int_{0}^{\infty}\frac{dk_1 dk_2 }{(2\pi)^2}e^{-k_1^2 t_{1,3}}e^{-k_2^2 t_{3,4}}e^{-k_1^2 t_{4,2}} G(n,k_1)G(n,-k_1) G(m-ik_1,-k_2)G(m+ik_1,k_2)
	\end{eqnarray}
	\begin{eqnarray}
	\langle g_n(t_1,t_2) g_m(t_3,t_4)\rangle_{V}= \int_0^\infty \frac{dk_1 dk_2dk_3}{(2\pi)^3} e^{-k_1^2t_{3,1}}e^{-k_2^2t_{1,4}}e^{-k_3^2t_{4,2}}  
	\nonumber \\ 
	G(m,k_1)G(n,-k_3)\frac{\Gamma(2ik_1)}{\Gamma(2ik_2)}\frac{\Gamma(-2ik_3)}{\Gamma(-2ik_2)}G(m+ik_2,k_3)G(n-ik_2,-k_1) \nonumber \\
	\int_{c_{\min(m,n)}-i\infty}^{c_{\min(m,n)}+i\infty}G(p+ik_2,k_1)\frac{\Gamma(n-p)}{\Gamma(n+p)}G(p,-k_3)\frac{\Gamma(m-p-ik_2)}{\Gamma(m+p+ik_2)}\frac{dp}{2\pi i} 
	\end{eqnarray}

	\section{Cancellation of infra-red singularities for 3rd and 4th time orderings}
	
	We need to calculate integrals like the one indicated in Eq.(9) of the main text:
	\begin{eqnarray}
	f(t_1-t_2)=\int_{t_3<t_4} (\langle g_n(t_1,t_2) g_m(t_3,t_4)\rangle-\langle g_n(t_1,t_2)\rangle \langle g_m(t_3,t_4)\rangle)
	\end{eqnarray} 
	We  introduce the following notations for the integrands corresponding to different variants of the time ordering:
	\begin{eqnarray}
	f_i(t_1-t_2)=\int_{T_i} dt_3 dt_4  \langle g_n(t_1,t_2) g_m(t_3,t_4)\rangle_i \nonumber \\
	f_Z = \int_{t_3<t_4, t_2<t_4,t_3<t_1} \langle g_n(t_1,t_2)\rangle \langle g_m(t_3,t_4)\rangle
	\label{f_def}
	\end{eqnarray}
	Here  $T_i$ is the area of integration which satisfies the $i$th order of times. Using these functions
	we can write: $f(t)=\sum_{i=II}^{V} f_i(t)-f_Z(t)$. The correction to the Green function can be expressed via $f(t)$
	as it is present in Eq.(10) of the main.  
	Note that  the functions  $f_{III}$, $f_{IV}$ and $f_{Z}$ are not well-defined since the integrals in
	Eq. (\ref{f_def}) diverge.  
	Fortunately, these divergencies  cancel each other.  To demonstrate with fact, we  write these function explicitly
	\begin{eqnarray}
	f_{III}(t_1-t_2)=\delta_{III} \langle g_n(t_1,t_2)\rangle=\int_{-\infty}^{t_2}dt_4 \int_{t_1}^{\infty} dt_3 \langle g_n(t_1,t_2) g_m(t_3,t_4)\rangle_{III}=\nonumber \\ \int_0^{\infty}\frac{dk_1}{2\pi}\int_0^{\infty}\frac{dk_2}{2\pi} \frac{e^{-k_2^2(t_1-t_2)}}{k_1^4} G(m,k_1)G(n+ik_1,k_2)G(n-ik_1,-k_2)G(m,-k_1)
	\end{eqnarray}
	and
	\begin{eqnarray}
	f_{IV}(t_1-t_2)=\delta_{IV} \langle g_n(t_1,t_2)\rangle=\int_{t_2}^{t_1}dt_3 \int_{t_2}^{t_3} dt_4 \langle g_n(t_1,t_2) g_m(t_3,t_4)\rangle_{IV}=\nonumber \\ 
	\iint_0^{\infty} \frac{dk_1}{2\pi}\frac{dk_2}{2\pi} \frac{e^{-k_2^2(t_1-t_2)}-e^{-k_1^2(t_1-t_2)}(1+(k_1^2-k_2^2)(t_1-t_2))}{(k_1^2-k_2^2)^2}
	G(n,k_1)G(m+ik_1,k_2)G(m-ik_1,-k_2)G(n,-k_1).
	\end{eqnarray}
	Finally,
	\begin{eqnarray}
	f_{Z}(t_1-t_2)=\delta_{Z} \langle g_n(t_1,t_2)\rangle =\int_{t_3>t_4,t_4<t_1,t_3>t_2} dt_3 dt_4 \langle g_n(t_1,t_2) \rangle \langle g_m(t_3,t_4)\rangle = \nonumber \\ \int_0^{\infty}\frac{dk_1}{2\pi}\int_0^{\infty}\frac{dk_2}{2\pi} \frac{e^{-k_1^2(t_1-t_2)}(1+k_2^2(t_1-t_2))}{k_2^4}G(n,k_1)G(n,-k_1)G(m,k_2)G(m,-k_2)
	\end{eqnarray}
	It is convenient to split $f_{Z}$ in two parts:
	\begin{eqnarray}
	f_{Z,III}(t)=\int_0^{\infty}\frac{dk_1}{2\pi}\int_0^{\infty}\frac{dk_2}{2\pi} \frac{e^{-k_2^2t}}{k_1^4}G(n,k_2)G(n,-k_2)G(m,k_1)G(m,-k_1) \\
	f_{Z,IV} = \int_0^{\infty}\frac{dk_1}{2\pi}\int_0^{\infty}\frac{dk_2}{2\pi} \frac{e^{-k_1^2t}t}{k_2^2}G(n,k_1)G(n,-k_1)G(m,k_2)G(m,-k_2)
	\end{eqnarray}
	The following combinations are free from divergencies upon integration:
	$\bar{f}_{III}(t)=f_{III}(t)-f_{Z,III}(t)$ and $\bar{f}_{IV}(t)=f_{IV}(t)-f_{Z,IV}(t)$. 
	In the next Section, we evaluate the asymptotic behaviour of the result of this integration.
	
	\section{Contribution from the regions III and IV.}
	
	In this Section we calculate contributions to the Green function correction coming from the time orderings 3 and 4.
	They can be represented explicitely as  some coefficient mutiplying $\langle g_n(t) \rangle$. To calculate it,
	we find the asymptotic behavior  of $\bar{f}_{III}(t)$ and $\bar{f}_{IV}(t)$ in the limit of long time $t$.
	We start from $\bar{f}_{III}(t)$. We  use here the fact  that for $t\gg 1$ one has $k_2\ll1$ and $k_1\sim 1$:
	\begin{eqnarray}
	\bar{f}_{III}(t)= \int_0^{\infty}\frac{dk_1}{2\pi}\int_0^{\infty}\frac{dk_2}{2\pi} \frac{e^{-k_2^2t}}{k_1^4} G(m,k_1)G(m,-k_1)(G(n+ik_1,k_2)G(n-ik_1,-k_2)-G(n,k_2)G(n,-k_2))\approx\nonumber \\
	\frac{\Gamma(n)^4}{2\sqrt{\pi}t^{\frac{3}{2}}}\int_0^{\infty}\frac{dk_1}{2\pi}\frac{G(m,k_1)G(m,-k_1)(\Gamma^2(n+ik_1)\Gamma^2(n-ik_1)-\Gamma^4(n))}{\Gamma(n)^4 k_1^4} \equiv C_{III}(n,m)\langle g_n(t)\rangle 
	\label{III}
	\end{eqnarray}
	
	To evaluate the contribution of the 4th time ordering it is convenient to split it into two parts.
	The first one is
	\begin{eqnarray}
	\label{OI}
	f_{IV,1}(t)= \int_0^{\infty}\frac{dk_1}{2\pi}\int_0^{\infty}\frac{dk_2}{2\pi} \frac{e^{-k_2^2t}-e^{-k_1^2t}}{(k_1^2-k_2^2)^2} G(n,k_1)G(m+ik_1,k_2)G(m-ik_1,-k_2)G(n,-k_1)= \nonumber \\
	\int_0^{\infty}\frac{dk_1}{2\pi}\int_0^{\infty}\frac{dk_2}{2\pi} \frac{e^{-k_2^2t}-e^{-k_1^2t}}{(k_1^2-k_2^2)^2}  \nonumber \\ \frac{\Gamma^2(n+ik_1)\Gamma^2(n-ik_1)\Gamma(m+ik_1+ik_2)\Gamma(m+ik_1-ik_2)\Gamma(m-ik_1+ik_2)\Gamma(m-ik_1-ik_2)}{\Gamma(2ik_1)\Gamma(-2ik_1)\Gamma(-2ik_2)\Gamma(-2ik_2)} \nonumber \\
	\end{eqnarray}
	We symmetrize it over interchange of $k_{1,2}$:
	\begin{eqnarray}
	f_{IV,1}(t)= \frac{1}{2}\int_0^{\infty}\frac{dk_1}{2\pi}\int_0^{\infty}\frac{dk_2}{2\pi} \frac{e^{-k_2^2t}-e^{-k_1^2t}}{(k_1^2-k_2^2)^2}  \nonumber \\ \frac{\Gamma^2(n+ik_1)\Gamma^2(n-ik_1)-\Gamma^2(n+ik_2)\Gamma^2(n-ik_2)}{\Gamma(-2ik_1)\Gamma(2ik_1)}G(m+ik_1,k_2)G(m-ik_1,-k_2). \nonumber \\
	\end{eqnarray}
	As a result:
	\begin{eqnarray}
	f_{IV,1}(t)= \int_0^{\infty}\frac{dk_1}{2\pi}\int_0^{\infty}\frac{dk_2}{2\pi} \frac{e^{-k_2^2t}}{(k_1^2-k_2^2)^2} \frac{\Gamma^2(n+ik_1)\Gamma^2(n-ik_1)-\Gamma^2(n+ik_2)\Gamma^2(n-ik_2)}{\Gamma(-2ik_1)\Gamma(2ik_1)}G(m+ik_1,k_2)G(m-ik_1,-k_2)\approx \nonumber \\ \frac{\Gamma(n)^4}{2\sqrt{\pi}t^{\frac{3}{2}}}\int_0^{\infty}\frac{dk_1}{2\pi} \frac{\Gamma^2(n+ik_1)\Gamma^2(n-ik_1)-\Gamma^4(n)}{k_1^4 \Gamma(n)^4}G(m,-k_1)G(m,k_1) = \nonumber \\
	= C_{III}(n,m) \langle g_n(t)\rangle \nonumber \\
	\end{eqnarray}
	To calculate the remaining terms from the 4th time ordering, we need to consider the following expression
	\begin{eqnarray}
	f_{IV,U}(t)=\int_0^{\infty}\frac{dk_1}{2\pi}\int_0^{\infty}\frac{dk_2}{2\pi} \frac{e^{-k_1^2t}t}{k_2^2-k_1^2}G(n,k_1)G(n,-k_1)G(m,k_2)G(m,-k_2)
	\end{eqnarray}
	Let us evaluate $f_{IV,2}-f_{IV,U}$:
	\begin{eqnarray}
	f_{IV,2}(t)-f_{IV,U}(t)=\int_0^{\infty}\frac{dk_1}{2\pi}\int_0^{\infty}\frac{dk_2}{2\pi} \frac{e^{-k_1^2t}t}{k_2^2-k_1^2}G(n,k_1)G(n,-k_1)(G(m+ik_1,k_2)G(m-ik_1,-k_2)-G(m,k_2)G(m,-k_2)) \nonumber \\
	\end{eqnarray}
	For $t\gg1$ one has $k_1\ll1$ and
	\begin{eqnarray}
	f_{IV,2}(t)-f_{IV,U}(t)=-\frac{\Gamma^4(n)}{2\sqrt{\pi}t^{\frac{3}{2}}}\int_0^{\infty}\frac{dk_2}{2\pi} \frac{3\Gamma^2(-ik_2+m)\Gamma^2(ik_2+m) (\psi^\prime(m-ik_2)+\psi^\prime(m+ik_2))\sinh(2k_2\pi)}{\pi k_2}= \nonumber \\
	\equiv C_{IV,1}(n,m) \langle g_n(t)\rangle,
	\label{IV1}
	\end{eqnarray}
	where $\psi$ is digamma function. The next step is to evaluate
	\begin{eqnarray}
	f_{IV,U}(t)-f_{Z,IV}(t)=\int_0^{\infty}\frac{dk_1}{2\pi}\int_0^{\infty}\frac{dk_2}{2\pi} \frac{e^{-k_1^2t}tk_1^2}{(k_2^2-k_1^2)k_2^2}G(n,k_1)G(n,-k_1)G(m,k_2)G(m,-k_2)
	\end{eqnarray}
	In this integral $k_{1}\ll1$   and
	\begin{eqnarray}
	f_{IV,U}(t)-f_{Z,IV}(t)\approx\int_0^{\infty}\frac{dk_1}{2\pi}\int_0^{\infty}\frac{dk_2}{2\pi} \frac{e^{-k_1^2t}4tk_1^4\Gamma^4(n)}{(k_2^2-k_1^2)k_2^2}G(m,k_2)G(m,-k_2)
	\end{eqnarray}
	To calculate the asymptotics, we add and subtract the following expression (below we will see that it is equal to zero):
	\begin{eqnarray}
	\delta=\int_0^{\infty}\frac{dk_1}{2\pi}\int_0^{\infty}\frac{dk_2}{2\pi} \frac{e^{-k_1^2t}4tk_1^4\Gamma^4(n)}{(k_2^2-k_1^2)k_2^2}\Gamma^4(m)4k_2^2.
	\end{eqnarray}
	This gives:
	\begin{eqnarray}
	f_{IV,U}(t)-f_{Z,IV}(t)-\delta=\int_0^{\infty}\frac{dk_1}{2\pi}\int_0^{\infty}\frac{dk_2}{2\pi} \frac{e^{-k_1^2t}4tk_1^4\Gamma^4(n)}{(k_2^2-k_1^2)k_2^2}(G(m,k_2)G(m,-k_2)-\Gamma^4(m)4k_2^2)
	\end{eqnarray}
	Here we can expand in $k_1$ and obtain
	\begin{eqnarray}
	f_{IV,U}(t)-f_{Z,IV}(t)-\delta=\frac{\Gamma^4(n)}{2\sqrt{\pi}t^{\frac{3}{2}}}\int_0^{\infty}\frac{dk_2}{2\pi} \frac{3}{2k_2^4}(G(m,k_2)G(m,-k_2)-\Gamma^4(m)4k_2^2) \equiv C_{IV,2}(n,m) \langle g_n(t)\rangle
	\label{IV2}
	\end{eqnarray}
	The last step to do is the calculation of $\delta$:
	\begin{eqnarray}
	\delta=16t\Gamma^4(n)\Gamma^4(m)\int_0^{\infty}\frac{dk_1}{2\pi}\int_0^{\infty}\frac{dk_2}{2\pi} \frac{e^{-k_1^2t}k_1^4}{k_2^2-k_1^2}=8t\Gamma^4(n)\Gamma^4(m)\int_0^{\infty}\frac{dk_1}{2\pi}\int_0^{\infty}\frac{dk_2}{2\pi} \frac{e^{-k_1^2t}k_1^4-e^{-k_2^2t}k_2^4}{k_2^2-k_1^2} = \nonumber \\8t\Gamma^4(n)\Gamma^4(m)\partial_{t}^2\int_0^{\infty}\frac{dk_1}{2\pi}\int_0^{\infty}\frac{dk_2}{2\pi} \frac{e^{-k_1^2t}-e^{-k_2^2t}}{k_2^2-k_1^2}=8t\Gamma^4(n)\Gamma^4(m)\partial_{t}^2\int_0^{\infty}\frac{dx}{2\pi}\int_0^{\infty}\frac{dy}{2\pi} \frac{e^{-x^2}-e^{-y^2}}{x^2-y^2}=0
	\end{eqnarray}
	Finally we have:
	\begin{eqnarray}
	\bar{f}_{IV}(t) \equiv f_{IV}(t) - f_Z(t) =(C_{III}(n,m)+C_{IV,1}(n,m)+C_{IV,2}(n,m))\langle g_n(t)\rangle
	\label{fIVbar}
	\end{eqnarray}
	where coefficients $C_i(n,m)$ are defined in Eqs.(\ref{III},\ref{IV1},\ref{IV2}).
	
	\section{Contributions from regions II and V and the final result}
	
	The time regions II and  V provides equal corrections to the Green function, so we will consider the region II only. 
	Here the correction to the Green function is
	\begin{eqnarray}
	f_{II}(t)=\int_0^t dt_3 \int_{-\infty}^0 \langle g_n(t,0)g_m(t_3,t_4)\rangle dt_3 dt_3 	= \nonumber \\\int_0^\infty \frac{dk_1 dk_2 dk_3}{(2\pi)^3} \frac{e^{-k_2^2t}-e^{-k_1^2 t}}{k_1^2-k_2^2} \frac{G(n,k_1)G(m,-k_3)G(n+ik_2,k_3)G(m-i k_2,-k_1)}{\Gamma(2i k_2)\Gamma(-2ik_2)} \nonumber \\
	\times G_{4,4}^{2,4}\left(\left|\begin{smallmatrix}
	1-ik_2-ik_1 & 1-ik_2+ik_1 & 1-ik_3 & 1+ik_3 \\ m & n-ik_2 & 1-m & 1-ik_2-n 
	\end{smallmatrix}\right|,1\right)
	\end{eqnarray} 
	Here $G^{2,4}_{4,4}$ is Meijer G-function. In the limit $t\rightarrow\infty$ we can obtain the following asymptotic formula
	for this function:
	\begin{equation}
	f_{II} (t) = \langle g_n(t)\rangle (C_{II,1}(n,m)+C_{II,2}(n,m))
	\label{fII}
	\end{equation}
	where coefficients $C_i(n,m)$ are given by
	\begin{eqnarray}
	C_{II,1}(n,m)=\int_0^\infty \frac{dk_2 dk_3 }{k_2 k_3 \pi^4 \Gamma(n)^2} 
	\Gamma(-ik_2+m)^2\Gamma(m-ik_3)\Gamma(m+ik_3)\Gamma(n+ik_2-ik_3)\Gamma(n+ik_2+ik_3)\nonumber \\
	\times
	G_{4,4}^{2,4}\left(\left|\begin{smallmatrix}
	1-ik_2 & 1-ik_2 & 1-ik_3 & 1+ik_3 \\ m & n-ik_2 & 1-m & 1-ik_2-n 
	\end{smallmatrix}\right|,1\right) \nonumber \\
	C_{II,2}(n,m)=\int_0^\infty \frac{dk_1 dk_3}{k_1 k_3 \pi^4 \Gamma(n)^4}\Gamma(m-ik_1)\Gamma(m+ik_1)\Gamma(m+ik_3)\Gamma(m-ik_3)\Gamma(n-ik_1)\Gamma(n+ik_1) \nonumber \\ \times  G_{4,4}^{2,4}\left(\left|\begin{smallmatrix}
	1-ik_1 & 1+ik_1 & 1-ik_3 & 1+ik_3 \\ m & n & 1-m & 1-n 
	\end{smallmatrix}\right|,1\right)
	\label{CII}
	\end{eqnarray}
	
	We combine now Eqs.(\ref{III},\ref{fIVbar},\ref{fII},\ref{CII}) to obtain the complete result for the relative correction to the Green function:
	\begin{eqnarray}
	\frac{\delta \langle G^{\frac{n}{\Delta}}\rangle }{\langle G^{\frac{n}{\Delta}}\rangle}= \frac{ N \Gamma^2 \Delta}{m}\frac{b^m}{\Gamma(2m)(2M)^{2m-2}}\left[2C_{III}(n,m)+C_{IV,1}(n,m)+C_{IV,2}(n,m)+2(C_{II,1}(n,m)+C_{II,2}(n,m))\right]
	\label{all}
	\end{eqnarray}
	Now we set $n=\frac{1}{4}$, $m=\frac{1}{2}$ and $\Delta = \frac{1}{4}$ in the above Eq.(\ref{all})
	and obtain the result for the first order correction to the Green function of the  $SYK_4$ model
	in presence of $SYK_2$ perturbation, as it is presented in Eq.(11) of the main text.
\end{widetext}

\end{document}